\documentclass[12pt]{article}
\usepackage{epsfig}
\usepackage{amsmath}
\usepackage{amssymb}
\usepackage{graphicx}
\usepackage{lscape}
\usepackage {subfigure}
\def\be{\begin{equation}}
\def\ee{\end{equation}}
\def\ba{\begin{array}}
\def\ea{\end{array}}
\def\beqn{\begin{eqnarray}}
\def\eeqn{\end{eqnarray}}

\def\bt{\begin{tabular}}
\def\et{\end{tabular}}
\def\bc{\begin{center}}
\def\ec{\end{center}}
\usepackage{epstopdf}
\usepackage{hyperref, cite}

\begin{document}
\title{Implication of rephasing invariant parameters on texture specific mass matrices}
\author{Madan Singh$^{*} $\\
\it Department of Physics, National Institute of Technology Kurukshetra,\\
\it Haryana,136119, India.\\
\it $^{*}$singhmadan179@gmail.com
}

\maketitle
\begin{abstract}
In the quark sector, Jarlskog rephasing invariant parameter $J_{CP}$ has important implications for the CP violation as well as phase structure of the quark mass matrices. In fact all CP violating effects in this sector are proportional to the magnitude of the imaginary part of $J_{CP}$ . With the observation of non-zero and large ”reactor mixing angle”, it is widely believed that CP might be violated in the leptonic sector. In this context, we have examined in detail
the relationship of CP odd invariants for texture specific mass matrices. In particular, we have
calculated the predicted ranges of Jarlskog rephasing invariant parameter and other weak basis
invariants for all the seven viable cases for texture two zero mass matrices.

\end{abstract}

\section{Intoduction}
At present, there is clear evidence that the Cabibbo-Kobayashi-Maskawa (CKM) matrix is complex, even if one allows for the presence of new physics in the $B_{d}$-$\overline{B}_{d}$ and $B_{s}$-$\overline{B}_{s}$  mixings. From a theoretical point of view, the complex phase in the CKM matrix may arise from complex Yukawa couplings and/or from a relative CP-violating phase in the vacuum expectation values (VEV) of Higgs fields. In either case, one expects an entirely analogous mechanism to arise in the lepton sector, leading to leptonic CP violation (LCPV).
The discovery of neutrino oscillations provides evidence for non-vanishing neutrino masses and leptonic mixing. Therefore, it is imperative to look for possible manifestations of CP violation in leptonic interactions. The recent measurement of reactor mixing angle $\theta_{13}$ and its subsequent refinements \cite{1,2,3,4}, along with the precision measurement of the solar and atmospheric mixing angles $\theta_{12}$ and $\theta_{23}$ as well as of the neutrino mass squared differences have given a new impetus to the neutrino oscillation phenomenology. The observation of non-zero value of $\theta_{13}$, on the one hand, restored the parallelism between quark and lepton mixing, while, it has triggered great deal of interest in the exploration of CP violation in the leptonic sector. The search for CP violation in the leptonic sector at low energies is one of the major challenges for experimental neutrino physics. Experiments with superbeams and neutrino beams from neutrino factories have the potential to measure either directly the Dirac phase $\delta$  through the observation of CP and T asymmetries or indirectly through neutrino oscillations. An alternative method is to measure the area of the unitarity triangles defined for the leptonic sector. In addition, the effects of Majorana type phases may arise in neutrinoless double beta decay ($0\nu \beta \beta$) processes. The observation of such processes would establish the Majorana nature of neutrinos and, possibly, provide some information on the Majorana CP phases. Thus, neutrino physics provides an invaluable tool for the investigation of leptonic CP violation at low energies apart from having profound implications for the physics of the early universe. In this direction, reconstruction of neutrino mass matrix is necessary as it encodes the information of CP violation, however in the present scenario of neutrino oscillation data it is impossible to fully construct neutrino mass matrix. Therefore, several proposals have been made in the literature to restrict the form of mass matrix \cite{5, 6, 7, 8, 9, 10, 11, 13}. However, texture zeros have been considered to be particularly successful in both flavor basis \cite{5, 8}  as well as non-flavor basis \cite{6, 7}. Particularly, in the flavor basis, many attempts have been made to explore the compatibility of texture zero mass matrices for Majorana neutrinos with the neutrino oscillation data \cite{5, 8, 12, 13}.  

In the flavor basis, where charged lepton mass matrix is diagonal, Frampton, Glashow and Marfatia (FGM) \cite{13}  have ruled out any possibility of survivabilty of neutrino mass matrix with three or more texture zeros, while in case of texture two-zero, only seven possibilities ($A_{1,2}, B_{1,2,3,4}, C$) out of total fifteen  are found to be viable with the experimental data.

\begin{equation}
A_{1}:\left(
\begin{array}{ccc}
0 & 0 & \times \\
0 & \times & \times\\
\times& \times & \times \\
\end{array}
\right),\; A_{2}:\left(
\begin{array}{ccc}
0& \times & 0 \\
\times & \times & \times\\
0& \times & \times \\
\end{array}
\right);\;
\end{equation}
\begin{equation}
\medskip
B_{1}:\left(
\begin{array}{ccc}
    \times& \times & 0 \\
  \times & 0 & \times\\
  0& \times & 0 \\
\end{array}
\right), B_{2}:\left(
\begin{array}{ccc}
    \times& 0 &\times \\
  0 & \times & \times\\
  \times& \times & 0 \\
\end{array}
\right), B_{3}:\left(
\begin{array}{ccc}
    \times& 0 &\times \\
  0 & 0 & \times\\
  \times& \times & \times \\
  \end{array}
\right),
  B_{4}:\left(
\begin{array}{ccc}
    \times& \times &0 \\
  \times & \times & \times\\
  0& \times & 0 \\
\end{array}
\right);
\end{equation}
\begin{equation}
C:\left(
\begin{array}{ccc}
    \times& \times & \times\\
  \times & 0 & \times\\
  \times& \times & 0 \\
\end{array}
\right),
\end{equation}

Here, the ‘$\times$’ denotes non-zero element.

The remaining eight classes belonging to categories D, E and F however, excludes the allowed experimental range of ratio of solar and atmospheric neutrino mass squared differences ($\delta m^{2}, \Delta m^{2}$) and hence ruled out with experimental data. In particular, the exact neutrino mass degeneracy ($m_{1} = m_{2} = m_{3}$) is predicted from three textures of Majorana neutrino mass matrix belonging to category F, which contradicts the result obtained from the solar neutrino oscillation experiments. In the light of non-zero and large measurement of $\theta_{13}$, several authors have recently attempted to carry out a detailed phenomenological implications of these textures for the effective Majorana mass, Dirac and Majorana CP-violating phases and neutrino masses \cite{8, 12}. In particular, it has been explicitly shown by M. Singh et. al. \cite{12} that the present refinements in data lead to the appreciable reduction in the available parameter space of the CP violating phases for viable texture two zero possibilities.

In the analysis of lepton flavor models, a useful approach when addressing the question of CP violation is the construction of the CP-odd weak basis (WB) invariants. Independent of the basis choice and phase convention, any of these quantities should vanish if CP is an exact symmetry of the theory. Thus, in CP violating theories which contain several phases, invariants constitute a powerful tool to investigate whether a particular texture zero model leads to leptonic CP violation at high and/or low energies.  In case of texture two zero, S Dev et. al \cite{14} have  presented the WB invariants in terms of elements of neutrino mass matrix for all the seven viable cases and consequently derived the CP invariant conditions for all viable cases. Similar analysis has been carried out by U. Sarkar and S. Singh \cite{15} for texture two zero neutrino mass matrices. In addition, the implications of CP-odd invariant parameters have been presented for other textures as well \cite{16}.
With the observation of non-zero and large "reactor mixing angle", it is widely believed that CP might be violated in the leptonic sector. Therefore, it becomes imperative to re-look at these CP–odd rephasing invariants in the context of texture two zero Majorana neutrino mass matrices. The purpose of present paper is to update the results of S. Dev et.al \cite{14} in the light of non-zero measurement of $\theta_{13}$ .

\section{Weak basis invariants and neutrino mass matrix}
In the flavor basis, where charged lepton mass matrix is diagonal, one can write the complex symmetric Majorana neutrino mass matrix as
\begin{equation}
M_{\nu}=\left(
\begin{array}{ccc}
   M_{ee}& M_{e \mu}& M_{e \tau} \\
  M_{e \mu} & M_{\mu \mu} & M_{\mu \tau}\\
  M_{e \tau}& M_{\mu \tau}& M_{\tau \tau} \\
  \end{array}
  \right),
  \end{equation}
  to denote the matrix elements of $M_{\nu}$. The above  matrix contains the information of neutrino mixing angle ($\theta_{12}, \theta_{23}, \theta_{13}$), three neutrino masses ($m_{1}, m_{2}, m_{3}$)  and three CP violating phases 
  ($\rho, \sigma$, $\delta$).\\
  For three generations of neutrino, the neutrino mixing matrix U can be parameterized by \cite{8}
 \begin{equation}
U=\left(
\begin{array}{ccc}
 c_{12}c_{13}& s_{12}c_{13}& s_{13} \\
-c_{12}s_{23}s_{13}-s_{12}c_{23}e^{-i\delta} & -s_{12}s_{23}s_{13}+c_{12}c_{23}e^{-i\delta} & s_{23}c_{13}\\
 -c_{12}c_{23}s_{13}+s_{12}s_{23}e^{-i\delta}& -s_{12}c_{23}s_{13}-c_{12}s_{23}e^{-i\delta}& c_{23}c_{13} \\
\end{array}
 \right),
\end{equation}
where $c_{ij} = cos \theta_{ij}, s_{ij}= sin
\theta_{ij}$ for i,j=1,2,3 and $\delta$ is the CP
violating phase.\\
The texture zeros are not weak basis (WB) invariants \cite{17}. This means that a given set of texture zeros which arise in a certain WB may not be present or may appear in different entries in another WB.  A large class of sets of leptonic texture zeros considered in the literature imply the vanishing of certain CP odd weak-basis invariants \cite{17}. Thus, we can recognize a lepton mass model in which the texture zeros are not explicitly present and which corresponds to a particular texture scheme in a certain WB. The relevance of CP odd WB invariants in the analysis of the texture zero ansatze is due to the fact that texture zeros lead to a decrease in the number of the independent CP violating phases. A minimum number of CP odd WB invariants can be found which will all vanish for the CP invariant mass matrices as a necessary and sufficient condition \cite{18}. 

In the present case, we address the question of finding CP odd WB invariants which would detect CP violation in the lepton sector. Firstly, we define the following three WB invariants, which must be non-zero for detecting CP violation in lepton sector \cite{19}

\begin{equation}
I_{1}=Im Det[H_{\nu }, H_{l}],
\end{equation}
\begin{equation}
I_{2}=Im Tr[H_{l }M_{\nu}M_{\nu}^{*}M_{\nu}H_{l}^{*}M_{\nu}^{*}, M_{\nu}^{*} ],
\end{equation}
\begin{equation}
I_{3}=Im Det[M_{\nu}^{*}H_{l}M_{\nu}, H_{l}^{*}]
\end{equation}

Here, $M_{l}$ and $M_{\nu}$ are the mass matrices for the charged leptons and the neutrinos, respectively, and $H_{l} = M_{l}^{\dagger} M_{l}$ and $H_{\nu} = M_{\nu}^{\dagger} M_{\nu}$ . The invariant $I_{1}$ was first proposed by Jarlskog \cite{20} as a rephasing invariant measure of Dirac type CP violation in the quark sector. It, also, describes the CP violation in the leptonic sector and is sensitive to the Dirac type CP violating phase. The invariants $I_{2}$ and $I_{3}$ were proposed by Branco, Lavoura and Rebelo \cite{21} as the WB invariant measures of Majorana type CP violation. The invariant $I_{3}$ was shown \cite{22} to have the special feature of being sensitive to Majorana type CP violating phase even in the limit of the exactly degenerate Majorana neutrinos.

As already mentioned in Ref. \cite{14}, invariant $I_{1}$ can be derived from $J_{CP}$. The Jarlskog rephasing invariant parameter $J_{CP}$ \cite{20} provides a measure of the Dirac-type CP violation and is defined as
\begin{equation}
Im[K_{ij}^{lm}]=J_{CP}\sum_{n}\epsilon_{lmn}
\sum_{k}\epsilon_{ijk},
\end{equation}\\ 
where $ K_{ij}^{lm} =U_{li}U_{lj}^{*} U_{mi}^{*}U_{mj}$. Here $U_{li}$ etc. represent the elements of U and $\varepsilon_{ilm}$, $\varepsilon_{ijk}$ denote the Levi-Civita symbols. Further, all the CP violating effects are directly proportional to $J_{CP}$ in the leptonic sector which is related to Dirac CP violating phase ($\delta$) through the relation
\begin{equation}
J_{CP}=s_{12} c_{12} s_{23}  c_{23} s_{13}  c_{13}^{2} \sin\delta.
\end{equation}
It may be emphasized that $J_{CP}$ is an 'invariant function' of mass matrices and is related to the mass matrices as
\begin{eqnarray}
&&
Det C=-2J_{CP}(m_{\tau}^{2}-m_{\mu}^{2})(m_{\mu}^{2}-m_{e}^{2})(m_{e}^{2}-m_{\tau}^{2})\nonumber  \\
&& \quad\quad \qquad(m_{\nu\tau}^{2}-m_{\nu\mu}^{2})(m_{\nu\mu}^{2}-m_{\nu e}^{2})(m_{\nu e}^{2}-m_{\nu\tau}^{2}),\qquad
\end{eqnarray} 
where
\begin{equation}
iC=[M_{l}M_{l}^{\dagger}, M_{\nu}M_{\nu}^{\dagger}].
\end{equation} \\
It may be mentioned that in the case of neutrino oscillation, the effect of Majorana phases does not appear in the U matrix. Therefore, it can be constructed  completely in case we know three mixing angles and the Dirac CP violating phase $\delta$. \\
In more concrete form,  $J_{CP}$  can be calculated in terms of  elements of neutrino mass matrix as \cite{14}
\begin{equation}
J_{CP}=\frac{Im(M_{ee}A_{ee}+M_{\mu \mu}A_{\mu \mu}+M_{\tau \tau}A_{\tau \tau})}{(m_{1}^{2}-m_{2}^{2})(m_{2}^{2}-m_{3}^{2})(m_{3}^{2}-m_{1}^{2})}.
\end{equation}

 \section{Inputs used in the present analysis}
 Before discussing the results of the analysis, we
summarize the experimental information about
various neutrino oscillation parameters. For both
normal mass ordering (NO) and inverted mass
ordering (IO), the best fit values and the latest
experimental constraints on neutrino parameters
at  3$\sigma$ confidence
level (CL), following Ref. \cite{23}, are
given in Table \ref{tab1}.
\begin{table*}[hb]
\begin{small}
\begin{center}
\begin{tabular}{|c|c|c|}
  \hline
  Parameter& Best Fit & 3$\sigma$ \\
  \hline
   $\delta m^{2}$ $[10^{-5}eV^{2}]$ & $7.50$& $7.03$ - $8.09$  \\
   \hline
   $|\Delta m^{2}_{31}|$ $[10^{-3}eV^{2}]$ (NO) & $2.52$ & $2.407$ - $2.643$ \\
   \hline
  $|\Delta m^{2}_{31}|$ $[10^{-3}eV^{2}]$ (IO) & $2.52$ &  $2.39$ - $2.63$ \\
  \hline
  $\theta_{12}$ & $33.56^{\circ}$ & $31.3^{\circ}$ - $35.99^{\circ}$\\
  \hline
  $ \theta_{23}$ (NO) & $41.6^{\circ}$  & $38.4^{\circ}$ - $52.8^{\circ}$ \\
  \hline
  $\theta_{23}$ (IO)& $50.0^{\circ}$ &  $38.8^{\circ}$ - $53.1^{\circ}$ \\
  \hline
  $\theta_{13}$ (NO) & $8.46^{\circ}$ &  $7.99^{\circ}$ - $8.90^{\circ}$ \\
  \hline
  $\theta_{13}$ (IO) & $8.49^{\circ}$ &  $8.03^{\circ}$ - $8.93^{\circ}$ \\
  \hline
  $\delta$ (NO) & $261^{\circ}$ & $0^{\circ}$ - $360^{\circ}$ \\
  \hline
  $\delta$ (IO) &$277^{\circ}$& $145^{\circ}$ - $391^{\circ}$ \\
\hline
\end{tabular}
\caption{  Best fit values and the latest
experimental constraints on neutrino parameters
at  3$\sigma$ confidence
level (CL) \cite{23} have been shown. NO (IO) refers to normal (inverted)
neutrino mass ordering.} \label{tab1}
\end{center}
\end{small}
\end{table*}\\
\section{Numerical analysis}
 For the present analysis, we have adopted the  method of random number generation for generating the data points for input parameters  within 3$\sigma$ error of neutrino oscillation data [Table \ref{tab1}].

\begin{figure}[ht]
\begin{center}
\subfigure[]{\includegraphics[width=0.35\columnwidth]{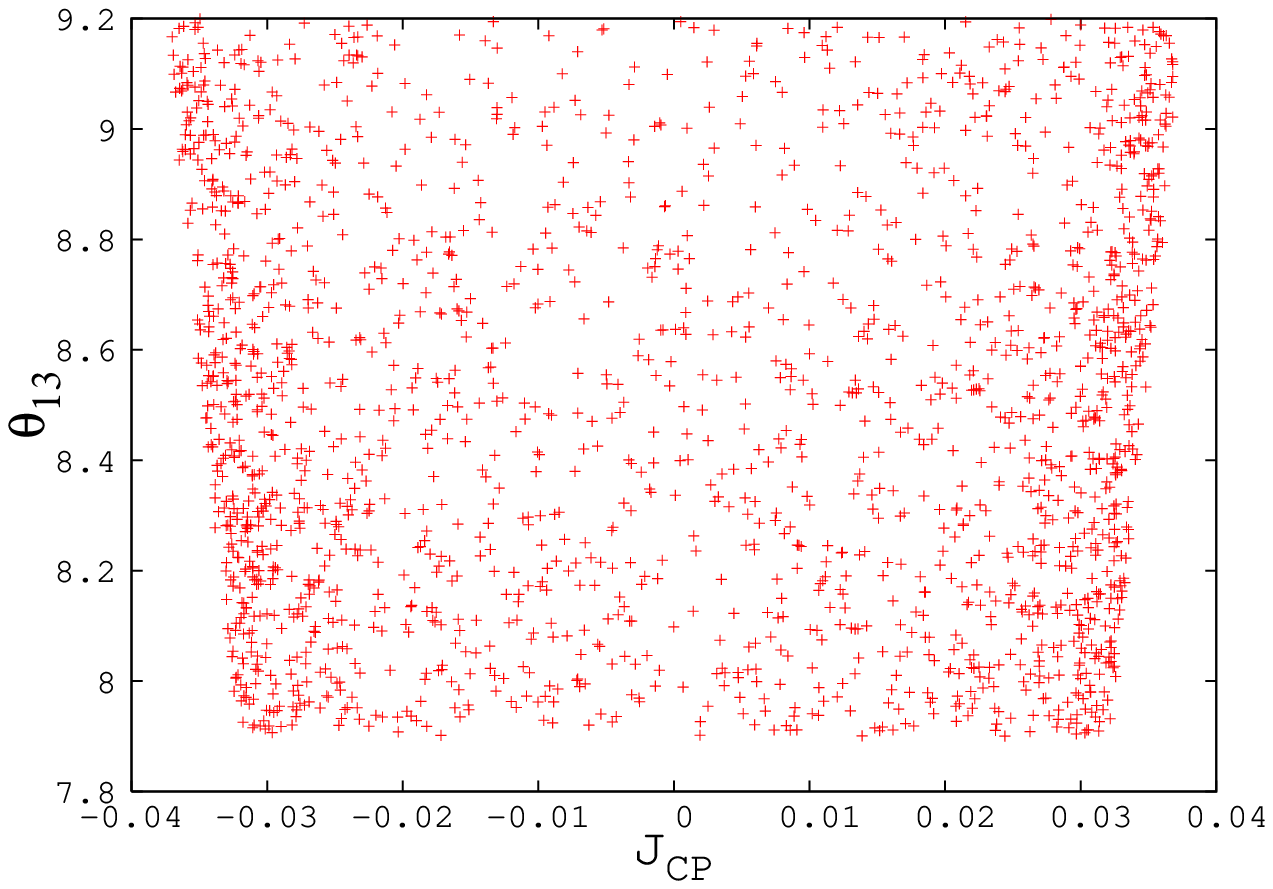}}
\subfigure[]{\includegraphics[width=0.35\columnwidth]{nA1}}\\
\caption{\label{fig1} Plot between Jarlskog rephasing invariant parameter $J_ {CP}$ and  reactor mixing angle  $\theta_{13}$ for  (a) case $A_{1}$ (b) case $A_{2}$. Angle $\theta_{13}$ is in degree. }
\end{center}
\end{figure}

\begin{table*}[ht]
\begin{small}
\begin{center}
\begin{footnotesize}
\begin{tabular}{|c|c|c|}
%  % after \\: \hline or \cline{col1-col2} \cline{col3-col4} ...
  \hline
Cases&Normal mass ordering
(NO) &Inverted mass ordering(IO) \\ 
\hline 
$A_{1}$  & $I_{1}=
($-$5.27-5.23)\times 10^{6}$ & $\times$  \\
 &$I_{2}=
($-$1.97-1.96)\times 10^{6}$ &$\times$ \\
&$I_{3}=
($-$5.88-5.92)\times 10^{23}$ & $\times$ \\
&$J_{CP}=$-$0.0376-0.0375$&$\times$ \\
\hline 
$A_{2}$ & $I_{1}=
($-$5.16-5.14)\times 10^{6}$ & $\times$  \\
 &$I_{2}=
($-$1.98-1.96)\times 10^{6}$ &$\times$ \\
&$I_{3}=
($-$2.0-2.0)\times 10^{21}$ & $\times$ \\
&$J_{CP}=$-$0.0377-0.0375$&$\times$\\
\hline
 $B_{1}$ & $I_{1}=
($-$1.26- $-$0.000476)\oplus (0.000427-1.24) \times 10^{6}$& $I_{1}=
($-$1.42-1.45)\times 10^{6}$  \\
 &$I_{2}=
($-$16.37-16.35)$&$I_{2}=
($-$11.53-11.47)$ \\
&$I_{3}=
($-$6.51-6.51)\times 10^{19}$ & $I_{3}=
($-$2.96-2.56)\times 10^{19}$ \\
&$J_{CP}=($-$0.00933-$-$0.000798)\oplus (0.000993-0.00922)$&$J_{CP}=$-$0.0107-0.0107 $\\
\hline
$B_{2}$ & $I_{1}=
($-$5.30-$-$1.636)\oplus (1.616-5.26) \times 10^{6}$& $I_{1}=
($-$5.30-$-$1.77)\oplus (1.699-5.31) \times 10^{6}$  \\
 &$I_{2}=
($-$20.01-20.19)$&$I_{2}=
($-$19.53-19.47)$ \\
&$I_{3}=
($-$1.96-1.96)\times 10^{23}$ & $I_{3}=
($-$1.59-1.15)\times 10^{17}$ \\
&$J_{CP}=($-$0.0376-$-$0.0284)\oplus (0.0287-0.0376)$&$J_{CP}=($-$0.0380-$-$0.0303)\oplus (0.0303-0.0380)$\\
\hline

$B_{3}$ & $I_{1}=
($-$5.44-$-$1.66)\oplus (1.66-5.44) \times 10^{6}$& $I_{1}=
($-$5.36-$-$1.63)\oplus (1.58-5.23) \times 10^{6}$  \\
 &$I_{2}=
($-$6.96-6.79)\times 10^{4}$&$I_{2}=
($-$9.12-9.21) \times 10^{4}$ \\
&$I_{3}=
($-$3.46-2.96)\times 10^{19}$ & $I_{3}=
($-$4.76-6.54)\times 10^{19}$ \\
&$J_{CP}=($-$0.0376-$-$0.0284)\oplus (0.0287-0.0376)$&$J_{CP}=($-$0.0377-$-$0.0287)\oplus (0.0289-0.0377)$\\
\hline

$B_{4}$ & $I_{1}=
($-$5.44-$-$1.66)\oplus (1.66-5.44) \times 10^{6}$& $I_{1}=
($-$5.30-$-$1.63)\oplus (1.61-5.26) \times 10^{6}$  \\
 &$I_{2}=
($-$15.3-15.3)$&$I_{2}=
($-$18.49-18.51)$ \\
&$I_{3}=
($-$7.19-3.57)\times 10^{17}$ & $I_{3}=
($-$1.301-1.303)\times 10^{17}$ \\
&$J_{CP}=($-$0.0376-$-$0.0284)\oplus (0.0287-0.0376)$&$J_{CP}=($-$0.0376-$-$0.0297)\oplus (0.0299-0.0374)$\\
\hline

$C$ & $I_{1}=
($-$1.66-1.06)\times 10^{6}$& $I_{1}=
($-$5.25-5.26) \times 10^{6}$  \\
 &$I_{2}=
($-$2.08-1.51)\times 10^{6}$&$I_{2}=
($-$1.05-1.05)\times 10^{5}$ \\
&$I_{3}=
($-$8.91-9.27)\times 10^{26}$ & $I_{3}=
($-$1.65-1.80)\times 10^{24}$ \\
&$J_{CP}=($-$0.0753-0.0754)$&$J_{CP}=($-$0.0371-0.0374)$\\
\hline
\end{tabular}
\caption{The predicted ranges of  $I_{1}  , I_{2}  , I_{3}$  and Jarlskog rephasing invariant parameter $J_{CP}$  for  the seven FGM cases of texture two zero has been presented. $I_{1}  , I_{2}  , I_{3}$  are in $(MeV/c)^{6} , (MeV/c)^{4}, (MeV/c)^{6}$,  respectively. The symbols - and $-$ denote mathematical minus and hyphen signs,  respectively. }\label{tab2}
\end{footnotesize}
\end{center}
\end{small}
\end{table*}

In the present work, we attempt to update the results of S. Dev et. al. \cite{14} with the present experimental data.  With the observation of non-zero measurement of reactor mixing angle, it  is  imperative to study the implication of the CP-odd invariants ($I_{1},I_{2},I_{3}$)  on texture two zero matrices in pursuit of CP violation in lepton sector.  Using the expressions of $I_{1}, I_{2}, I_{3}$, given in Ref. \cite{14},  we have calculated the ranges of  $I_{1}, I_{2}$ and $I_{3}$  for all the seven viable cases of texture two zero mass matrix for both normal (NO) and inverted (IO) mass ordering. 

\begin{figure}[ht]
\begin{center}
\subfigure[]{\includegraphics[width=0.45\columnwidth]{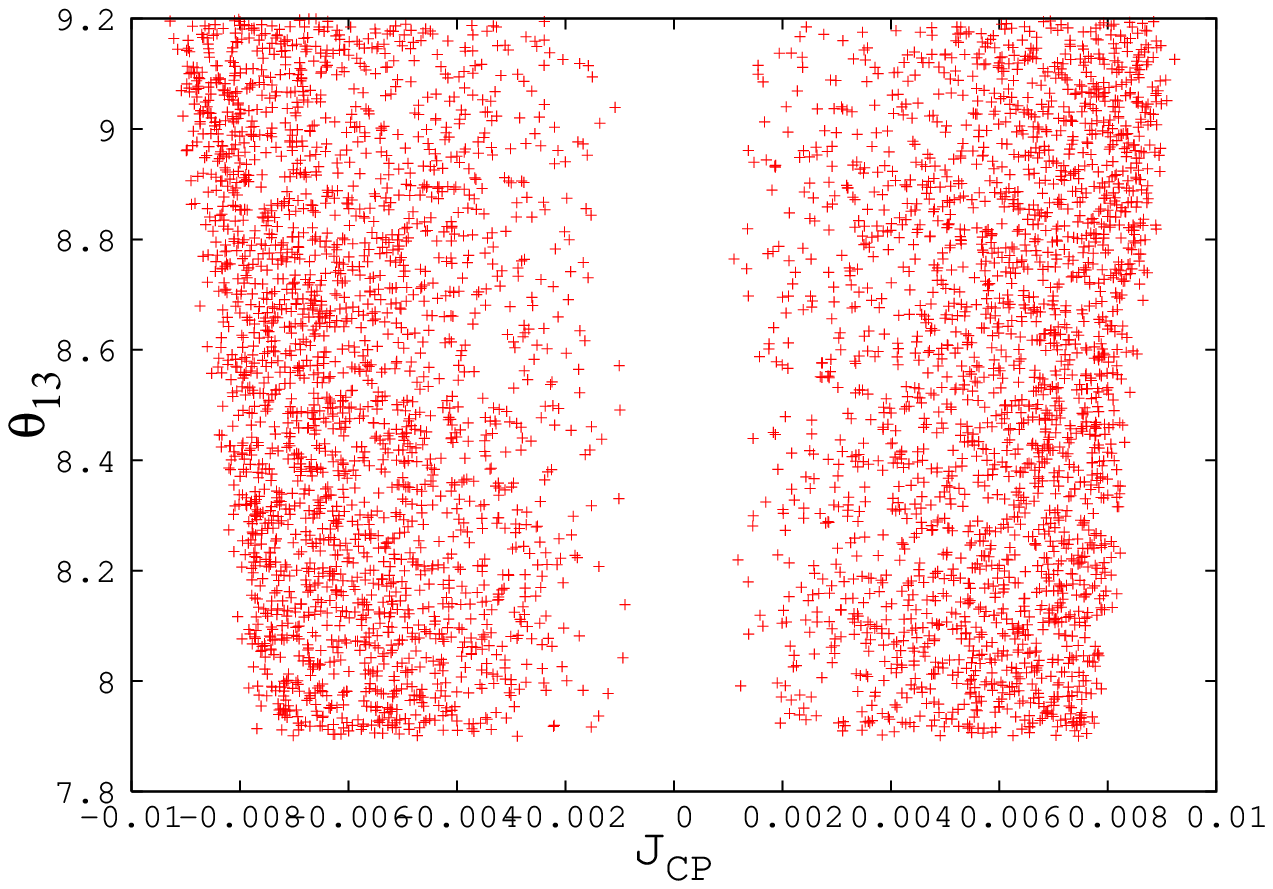}}
\subfigure[]{\includegraphics[width=0.45\columnwidth]{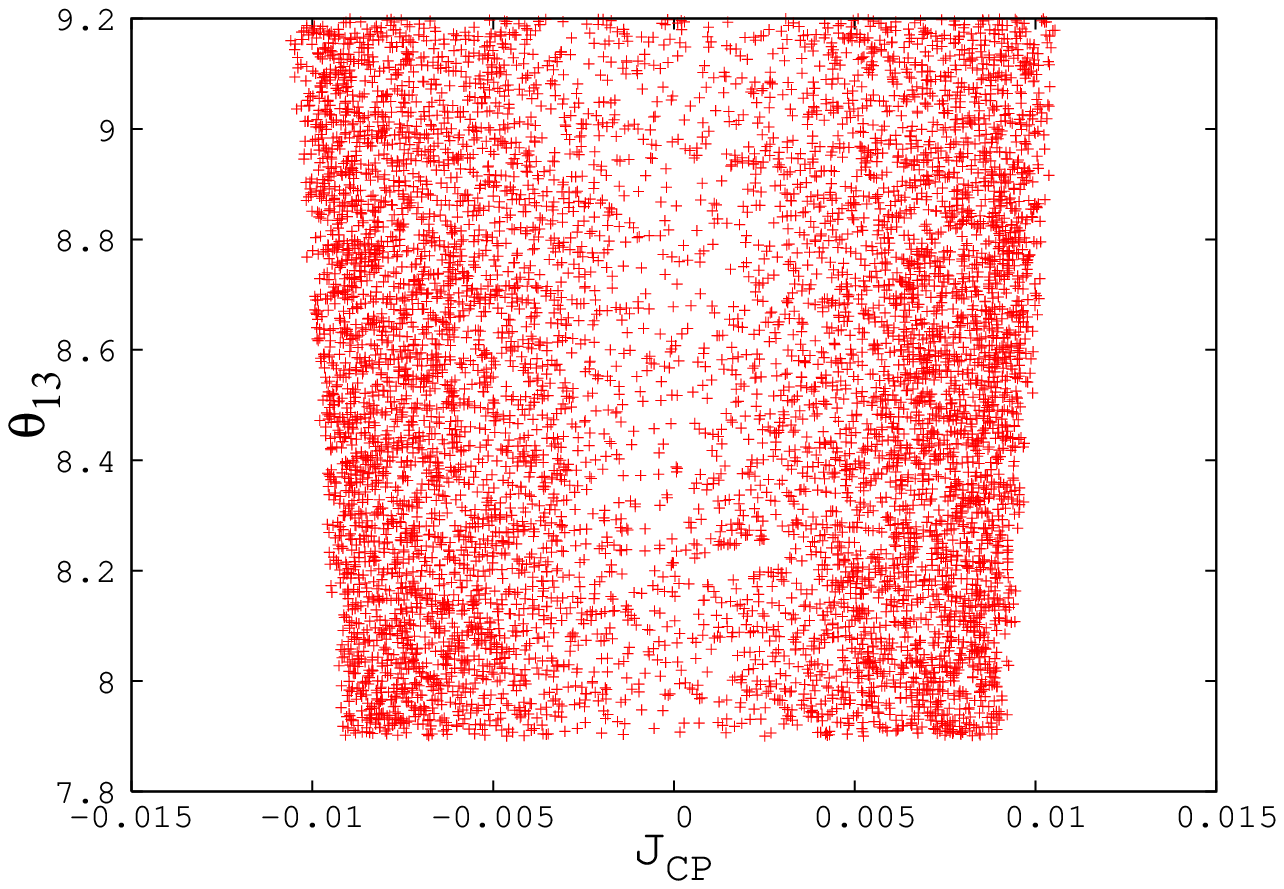}}\\
\caption{\label{fig2} Plot between Jarlskog rephasing invariant parameter $J_ {CP}$ and  reactor mixing angle  $\theta_{13}$ for  case $B_{1}$ (a) NO (b) IO. Angle $\theta_{13}$ is in degree. }
\end{center}
\end{figure}
As shown in Ref. \cite{8}, cases $A_{1}$ and $A_{2}$ predict NO, however IO remain ruled out at 3$\sigma$ CL, while cases $B_{1,2,3,4}$ and C predict both NO and IO at same CL. In addition cases $B_{1,2,3,4}$ and C exhibit quasi-degenerate neutrino mass spectrum.  Cases $B_{1,2,3,4}$ also predict nearly maximal CP violation.

\begin{figure}[ht]
\begin{center}
\subfigure[]{\includegraphics[width=0.35\columnwidth]{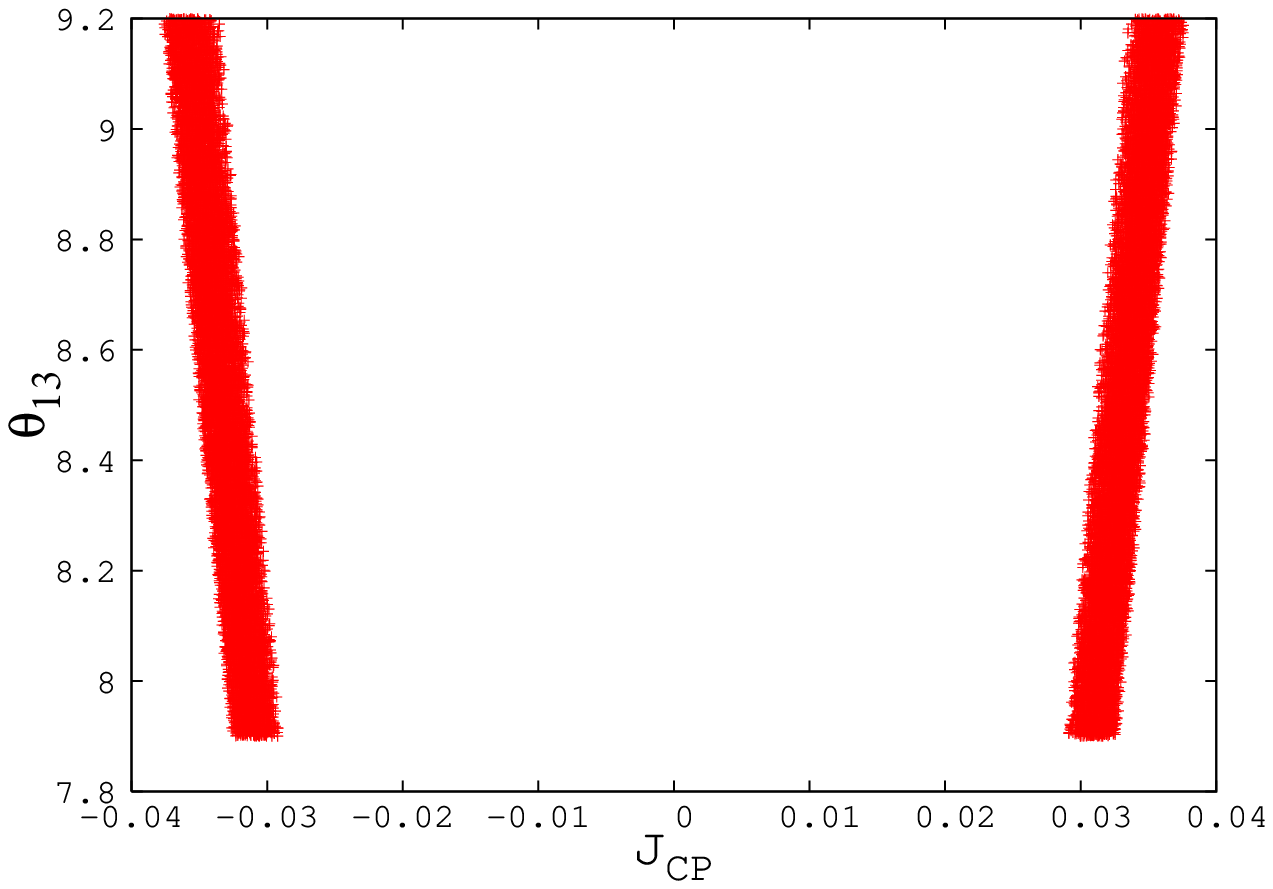}}
\subfigure[]{\includegraphics[width=0.35\columnwidth]{B2j.eps}}\\
\caption{\label{fig3} Plot between Jarlskog rephasing invariant parameter $J_ {CP}$ and  reactor mixing angle  $\theta_{13}$ for  case $B_{2}$ (a) NO (b) IO. Angle $\theta_{13}$ is in degree. }
\end{center}
\end{figure}

In Fig.\ref{fig1}, it is shown that for cases $A_{1,2}$, the $J_{CP}$  ranges from  $-$0.0377 to 0.0377, which implies that these cases strongly predict both the possibilities of CP violation and conservation.  From Table \ref{tab2}, the calculated range of $I_{1}$ for these cases further affirms these predictions since $I_{1}$ is directly proportional to $J_{CP}$  \cite{14}. Further, cases $B_{1}$ (IO) and C (NO and IO) hold more or less similar predictions for invariant $I_{1}$ as for cases $A_{1, 2}$ [Figs. \ref{fig2}(b), \ref{fig6}].  
                                     
\begin{figure}[ht]
\begin{center}
\subfigure[]{\includegraphics[width=0.35\columnwidth]{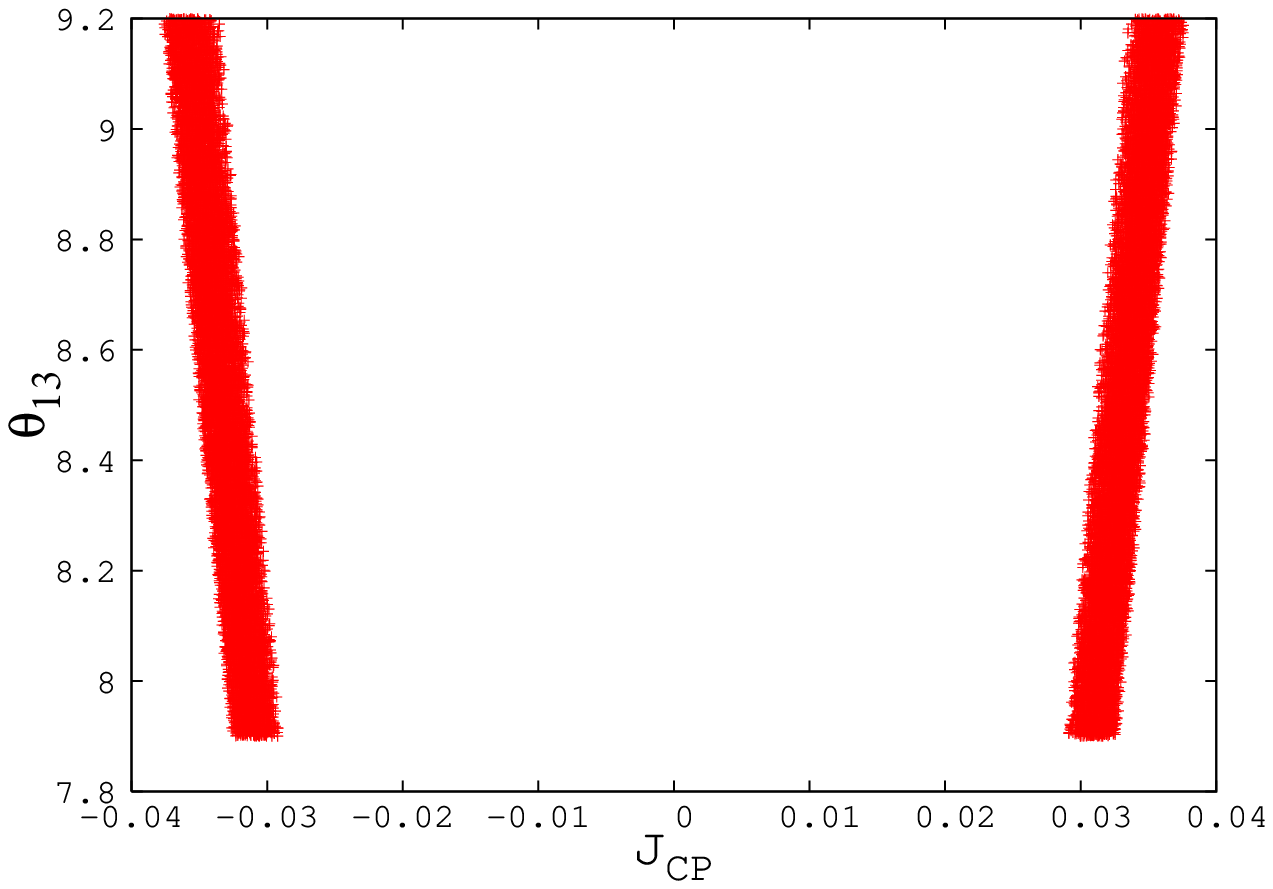}}
\subfigure[]{\includegraphics[width=0.35\columnwidth]{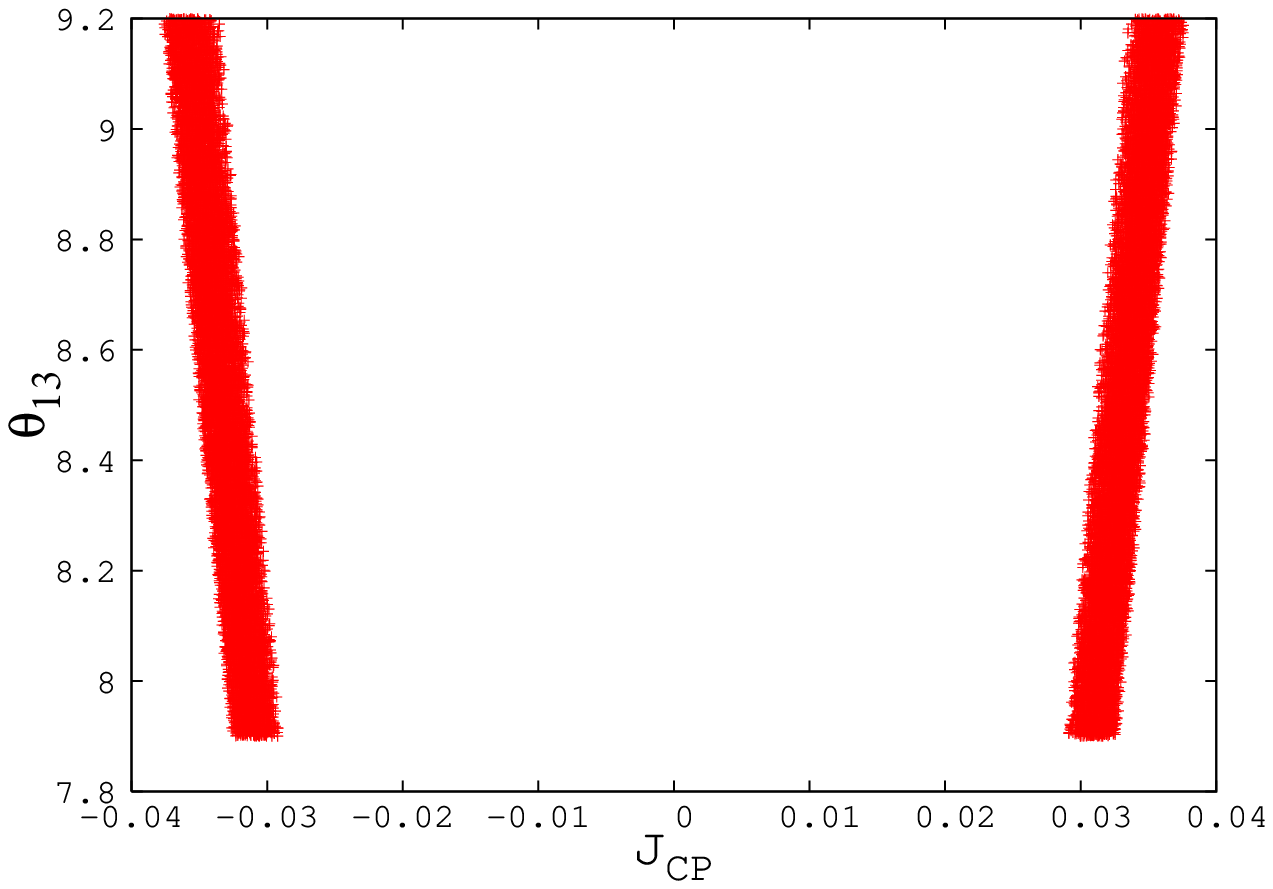}}\\
\caption{\label{fig4} Plot between Jarlskog rephasing invariant parameter $J_ {CP}$ and  reactor mixing angle  $\theta_{13}$ for  case $B_{3}$  (a) NO (b) IO. Angle $\theta_{13}$ is in degree. }
\end{center}
\end{figure}
	                                          
	  \begin{figure}[ht]
\begin{center}
\subfigure[]{\includegraphics[width=0.35\columnwidth]{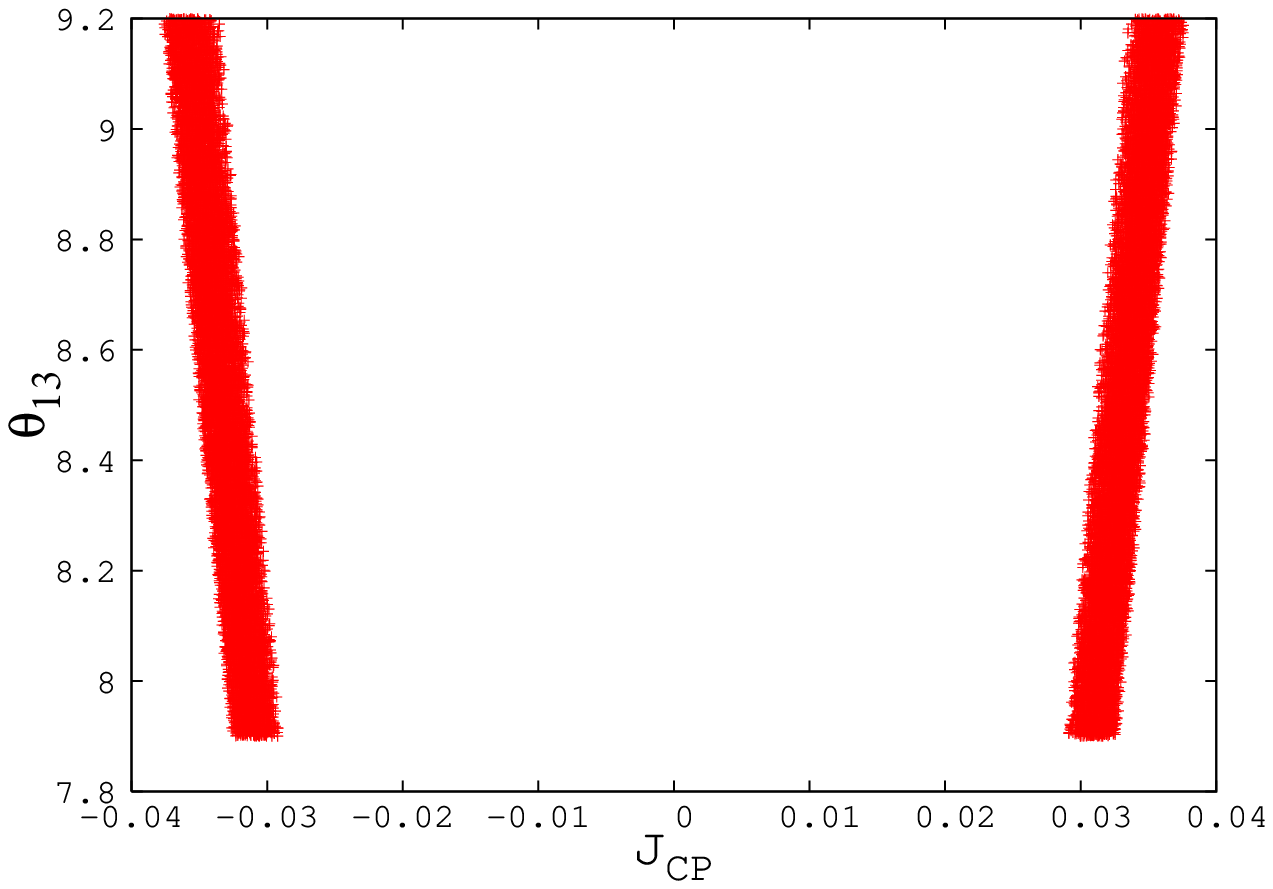}}
\subfigure[]{\includegraphics[width=0.35\columnwidth]{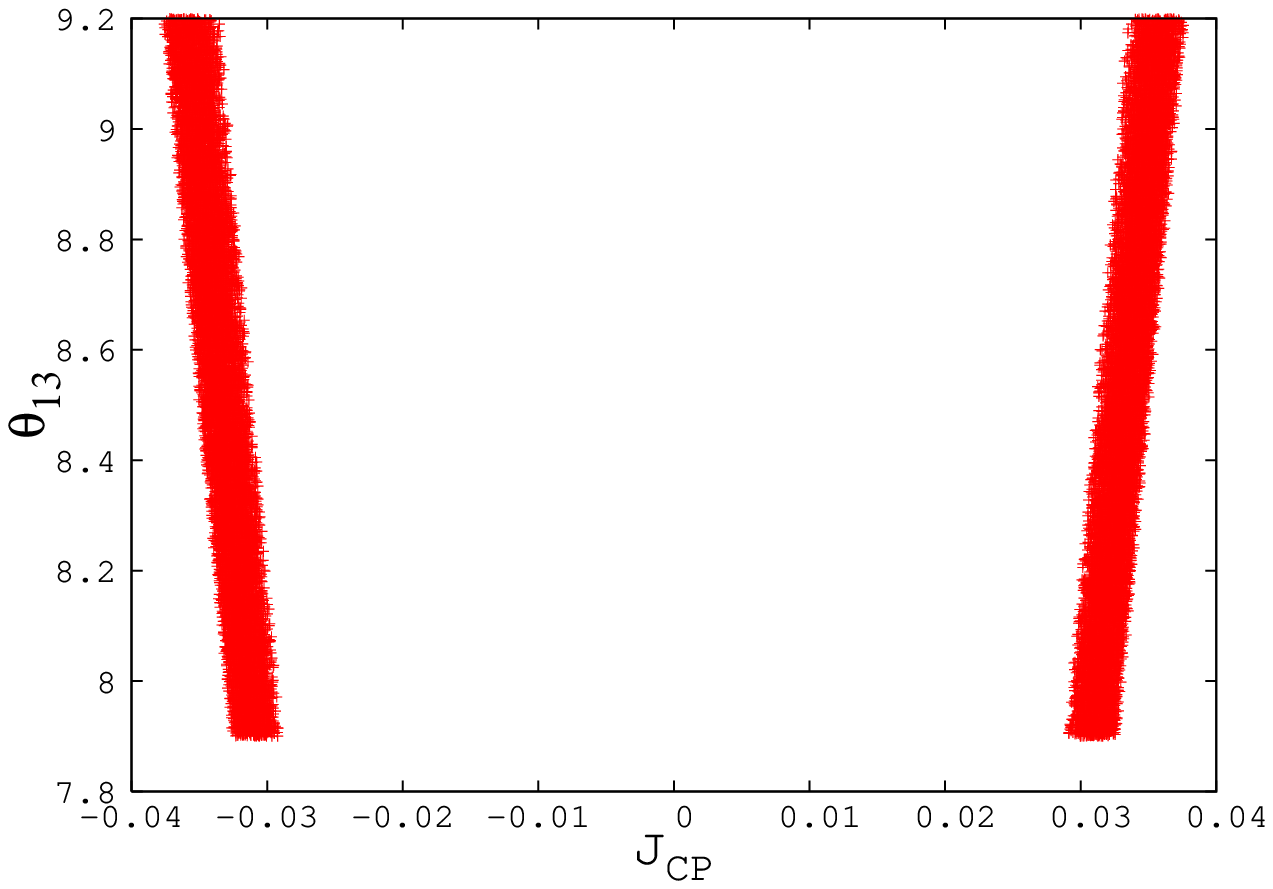}}\\
\caption{\label{fig5} Plot between Jarlskog rephasing invariant parameter $J_ {CP}$ and  reactor mixing angle  $\theta_{13}$ for  case $B_{4}$  (a) NO (b) IO. Angle $\theta_{13}$ is in degree. }
\end{center}
\end{figure}
	                                          
For remaining cases $B_{1,2,3,4}$, we have presented the correlation plot for $J_ {CP}$ and $\theta_{13}$ at 3$ \sigma$ CL. As explicitly shown in Fig. \ref{fig3}, Fig. \ref{fig4} and Fig. \ref{fig5}, $J_ {CP}$  is non zero and hence $I_{1}$ is non-zero, implying that CP is necessarily violating for these cases in  lepton number conserving processes (LNV). Moreover, $J_ {CP}$ is nearly maximal,  which further implies that CP is maximally violated for these cases [Fig.\ref{fig3}, Fig.\ref{fig4}, Fig.\ref{fig5}]. 
In Ref. \cite{24}, using the $\chi^{2}$ analysis, the maximum allowed CP violation,   $J_{CP}^{max}= 0.0329 \pm 0.0009 (\pm 0.0027)$ is determined at 1$\sigma$ (3$\sigma$) for both orderings. The preference of the present data for non-zero $\delta$ implies a best fit  $J_{CP}^{max}= $ $-$0.032, which is favored over CP conservation at the $\sim 1.2 \sigma$ level. Therefore at present, our results are in tune with these results, however only the future refinement in experimental data can shed some light on these predictions, which may either rule out or put some stringent constraint on $J_ {CP}$.

      \begin{figure}[ht]
\begin{center}
\subfigure[]{\includegraphics[width=0.35\columnwidth]{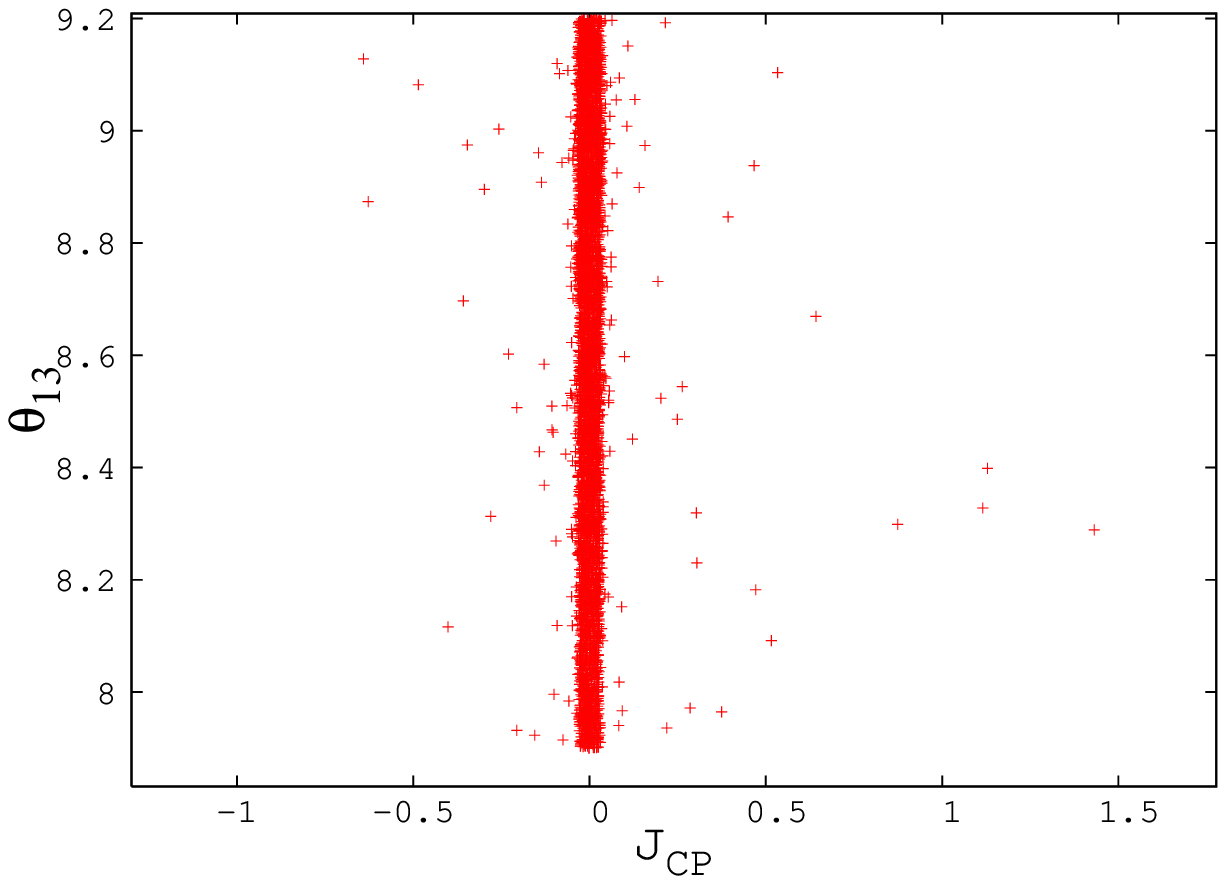}}
\subfigure[]{\includegraphics[width=0.35\columnwidth]{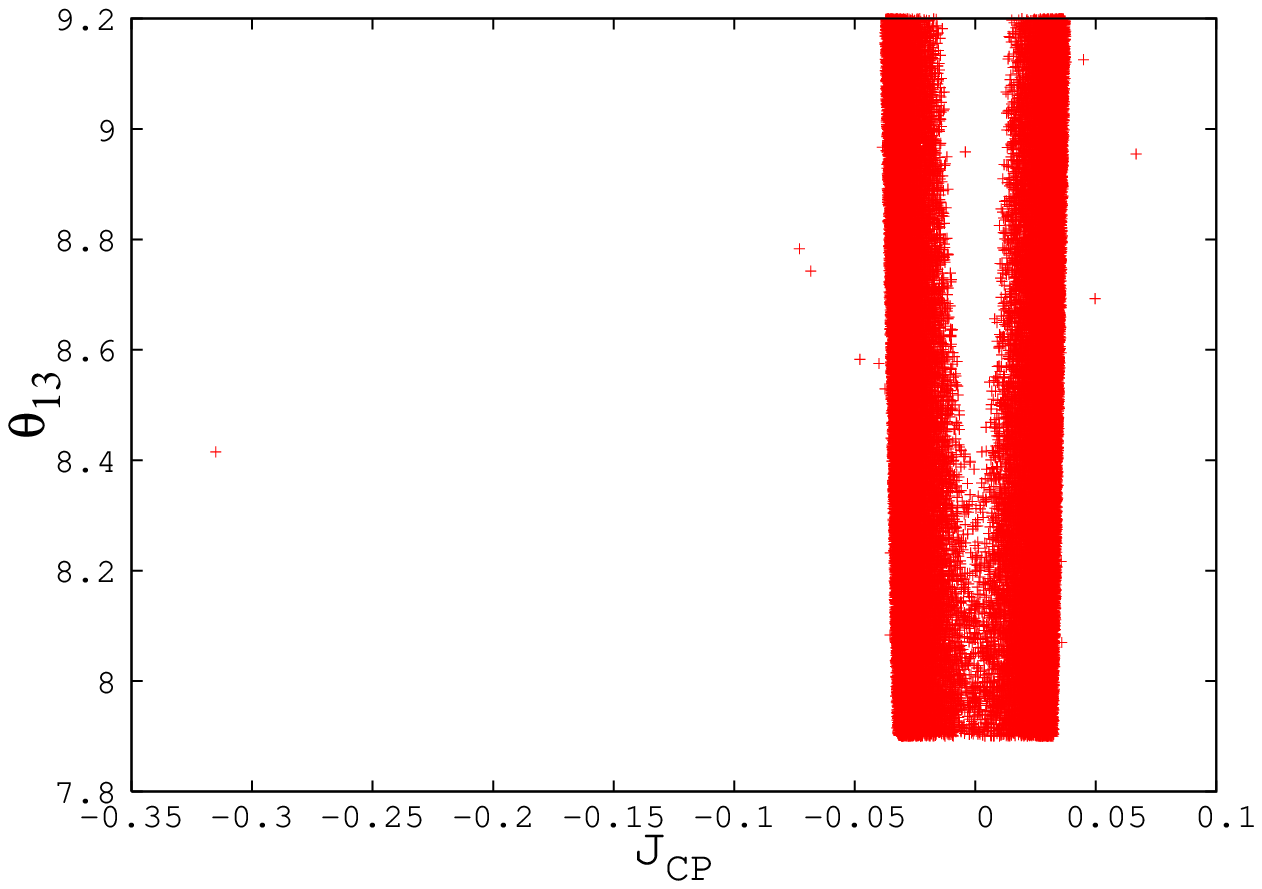}}\\
\caption{\label{fig6} Plot between Jarlskog rephasing invariant parameter $J_ {CP}$ and  reactor mixing angle  $\theta_{13}$ for  case $C$  (a) NO (b) IO. Angle $\theta_{13}$ is in degree. }
\end{center}
\end{figure}                      
As far as rephasing invariants $I_{2}$ and $I_{3}$  are concerned, as already stated that these invariants are sensitive to the measurement of Majorana type phases. In the limit of degenerate neutrino masses, $I_{3}$ have been found to be particularly sensitive to Majorana type CP violating phases \cite{22}. From Table \ref{tab2}, it is clear that for five cases ($B_{1,2,3,4}$ and C)  predicting the degenerate neutrino masses spectrum, $I_{3}$ is found to be of the order of O($10^{17}$-$10^{26}$), which is extremely large compared to the WB invariant $I_{2}$. Therefore at present, cases $B_{1,2,3,4}$ and C appear to be a strong candidate to predict CP violation in lepton number violating (LNV) processes through WB invariant $I_{3}$. However, only observations of double beta decay \cite{25, 26} would imply lepton number violation and Majorana nature of neutrinos. 

\section{Summary and conclusions}

In the light of non-zero measurement of  $\theta_{13}$ , we have updated the results of S. Dev et. al. \cite{14}  by calculating the ranges of WB invariants $I_{1}, I_{2}, I_{3}$ for viable texture two cases. It has been found that invariant $I_{1}$, which is sensitive to Dirac CP violating phase $\delta$ covers reasonably a large parameter space including vanishing value for cases A , $B_{1}$(IO) and C,   respectively. Hence, one can conclude that cases A,  $ B_{1}$(IO) and C  predict both the possibilities viz. CP violation as well as  CP conservation. In this regard, we have explicitly shown the correlation plots between $J_{CP}$ and reactor mixing angle $\theta_{13}$. On the other hand, for remaining cases $B_{1}$(NO), $B_{2}, B_{3}, B_{4},  I_{1}$ is strictly non-zero and hence, $J_{CP}$  is necessarily CP violated as well as indicates the maximal CP violation. The other invariant $I_{3}$ is also found to be very large in magnitude (including the vanishing value) for all the viable cases. The significance of $I_{3}$ is found to be prominent in case of $B_{1, 2, 3, 4}$ and C cases. In addition, parameter $I_{2}$ which is also sensitive to Majorana phases found to be negligibly small for $B_{1,2,3,4}$ cases. At present, these indications predict  CP violation and signal towards the possible CP violations in both LNC and LNV process, respectively, however only the future double beta decay experiments (LNV process) and neutrino oscillations (LNC processes) as well as the cosmological data could throw some light on these predictions.

\section*{Conflicts of Interest}
The authors declare that there are no conflicts of interest regarding the publication of this paper.

\section*{Acknowledgment}

The author would like to thank the Director, National Institute of Technology Kurukshetra, for providing necessary facilities to work.  \\

\newpage

\end{document}